\documentstyle[prc,preprint,aps]{revtex}
\begin{document}
\draft
\title{Strangeness Form Factors of the Proton in the Chiral Quark Model}
\author{L. Hannelius and D.O. Riska}
\address{Department of Physics, 00014 University of Helsinki, Finland}
\maketitle
\begin{abstract}
The chiral quark model describes the strangeness components 
of the light quarks as fluctuations into strange mesons and quarks. The
single strange pseudoscalar and vector meson loop fluctuations of the
constituent $u$- and $d$-quarks give rise to only very small 
strangeness form factors for the proton. This result is in line
with recent experimental results, given their wide uncertainty range.
\end{abstract}
\section{Introduction}
The HAPPEX experiment \cite{Happex} shows the combination 
$G_E^s+0.39 G_M^s$ of the strange charge and magnetic form factors 
of the proton at $Q^2=0.48$ (GeV/$c)^2$ to be consistent with 
$0 \,(0.023\pm0.034\pm0.022\pm0.026)$. Similarly the SAMPLE experiment
\cite{Sample1} shows that $G_M^s$ at $Q^2=0.1$ (GeV/$c)^2$ is consistent
with 0, modulo uncertainties in the calculated value of the weak axial
form factor of the nucleon~\cite{Sample2}. The experimental result that
the strangeness form factors of the proton are small may be used to
constrain or test theoretical models for nucleon structure, as the
theoretical predictions for these observables have covered a fairly wide
range~\cite{Jaffe,MuBu,Weigel,Hammer,Barz,Barz2}. 

A calculation of the strangeness form factors of the proton based
on the chiral quark model is reported here. The approach considers the
strangeness component of the proton as loop fluctuations, with
intermediate strange mesons and $s$-quarks, of the
constituent quarks that form the proton. The constituent quark model
approach represents an
alternative to the hadronic approach, in which the strangeness components
are considered as fluctuations of the nucleon into intermediate strange
mesons and hyperons. The chiral quark model approach brings the
advantages of much smaller coupling constants and consequently the
possibility of a
converging loop expansion that takes 
all baryonic intermediate
states into account. Moreover, as the amplitudes of the loops mainly scale with the
inverse squared mass of the intermediate meson, heavy meson
contributions are suppressed. Examples of this
are the desirably small meson loop contributions to the neutron charge
radius~\cite{Gloz} and the
recent demonstration that the strange meson loop contributions to the
proton give rise to but a very small strangeness magnetic
moment~\cite{Hannelius}.

It is shown here that the kaon and $K^*$ loop fluctuations of the
light quarks, that are illustrated in Fig.~1, and which
lead to but a small value for the strangeness magnetic moment
of the proton, also lead to strange form factors
of the proton with very small magnitudes.
The results are found to be fairly insensitive to the value of the
cut-off scale for the loop integrals, provided that this is taken to be
about 1 GeV or larger, ie.~values commensurate with the chiral symmetry
restoration scale $\sim 4\pi f\pi \sim 1.2$ GeV, at which scale
the pseudoscalar mesons are expected to decouple from the constituent quarks.

The magnitude of the strange loop contributions to the strangeness form
factors of the proton is of the order $(g^2/4\pi^2)(m_q^2/(m_q^2+m_M^2))$,
where $g$ is the meson-quark coupling, and $m_q$ and $m_M$ are the
masses of the light constituent quarks and strange mesons respectively.
As $g^2/4\pi \sim 0.7$ for $K$ and $K^*$ mesons, it follows with
$m_q\sim 300$ MeV that the loop amplitude in the case of kaons is
expected to only be about $\sim 0.06$, and smaller still
in the case of $K^*$
mesons. In comparison the expected magnitude of a typical loop
amplitude in the
case of the hadronic approach, where $g^2/4\pi \simeq 10$ and $m_q$
is replaced by the proton mass, is more than an order of
magnitude greater. This is also revealed by a comparison of the
calculated values for the strangeness magnetic moment of the proton in
Refs.~\cite{Barz} and~\cite{Hannelius}. It suggests that a small net loop
contribution in the hadronic approach only can result as a consequence of
strong cancellations between several large amplitudes unless strong
cut-offs are invoked.

This paper falls into 5 sections. In section 2 the contribution of the
strange loop amplitudes to the proton form factors, that contain
intermediate kaons are derived. The corresponding results for the loop
amplitudes that involve intermediate $K^*$ mesons are derived in section
3. The contribution from loop fluctuations with $K^*\rightarrow K\gamma$
vertices is derived in section 4.
The numerical results for the strangeness form factors of the proton
are calculated in section 5. 

\section{Kaon loop contributions}

The strangeness form factors of the nucleon are defined as the invariant
coefficients of the matrix elements of the operators $\bar s\gamma_\mu s$
in the proton. In standard notation the strangeness current is 
\begin{equation}
\langle p'|j_\mu^s(0)|p\rangle =
i\bar u(p')\Bigl[F_1^s(Q^2)\gamma_\mu
-F_2^s(Q^2)\frac{\sigma_{\mu\nu}q_\nu}{2m_N}\Bigr]u(p).
\eqnum{2.1}
\end{equation}
Here $q=p'-p$, and $Q^2=q^2=\bbox{q}^2-q_0^2>0$, and $m_N$ is the proton mass.
The two form factors are 
calculated here in the constituent quark model from the strangeness loop
fluctuations illustrated by the Feynman diagrams in Fig.~1, where the
meson lines represents $K$ and $K^*$ mesons.

Consider first the kaon loop diagrams. To calculate these, we consider
the kaon-quark pseudovector coupling:
\begin{equation}
{\cal L}=i\frac{f_{Kqs}}{m_K}\bar\psi
\gamma_5\gamma_\mu\sum_{a=4}^{7}\lambda^a\partial_\mu K^a\psi. \eqnum{2.2}
\end{equation}
The pseudovector kaon-quark coupling constant is
obtained as~\cite{Hannelius}
\begin{equation}
f_{Kqs}=\frac{g_A^a}{2}\frac{m_K}{f_K}, \eqnum{2.3}
\end{equation}
where $g_A^q=0.87$ for quarks~\cite{Wein,Dicus} and $f_K$ is the kaon
decay constant $(f_K=113$ MeV). The numerical value for
$f_{Kqs}$ is then $f_{Kqs}=1.9$.

The kaon and strange quark current density operators have the form
\begin{eqnarray}
j_\mu =&&i e\{\partial_\mu K^\dagger K+\text{h.c.}\}, \eqnum{2.4a}\\
j_\mu =&& -i\frac{e}{3}\bar \psi_s\gamma_\mu\psi_s. \eqnum{2.4b}
\end{eqnarray}
The standard convention on the strangeness form factors assigns the
$s$-quark a strangeness charge of $+1$ and the kaon, which contains an
$\bar s$-quark, a strangeness charge of $-1$. In the calculation of the
loop amplitudes the $s$-quark current
(2.4b) should therefore be multiplied by $-3$ and the kaon current
(2.4a) by $-1$.

The pseudovector coupling term (2.2) requires introduction of a contact
coupling term for current conservation. This contact current term gives
rise to two seagull diagrams, which exactly cancel the corresponding
seagull diagrams, which arise in the evaluation of the amplitudes of the
loop diagrams in Fig.~1, upon application of the Dirac equation for the
external quarks. The remaining loop amplitudes are equivalent to those,
which arise if the loop amplitudes are calculated using the pseudoscalar
coupling
\begin{equation}
{\cal L}_{Kqs}=ig_{Kqs}\bar \psi\gamma_5\sum_{a=4}^{7}\lambda^aK^a
\psi, \eqnum{2.5}
\end{equation}
where the pseudoscalar coupling constant $g_{Kqs}$ is defined as
\begin{equation}
g_{Kqs}=\frac{m_q+m_s}{m_K}f_{Kqs}. \eqnum{2.6}
\end{equation}
In these expressions $m_q$ represents the constituent mass of the light
flavor quarks $(u,d)$ and $m_s$ represents the strange quark mass. For
these masses we shall use the values $m_q=340$ MeV and $m_s=460$ MeV
respectively~\cite{Hannelius}. With these mass values we obtain the
value $g_{Kqs}^2/4\pi=0.75$ for the (squared) kaon-quark
pseudoscalar coupling constant.

The kaon loop contributions to the Dirac form factors $F_1^s$ of the
quarks are logarithmically divergent. We regularize these loop amplitudes
by cutting off the loop momentum integrals smoothly at the chiral
restoration scale $\Lambda_\chi=4\pi f_\pi=1.2$ GeV. The cut-off is
implemented by replacing the meson propagator $v(k^2)=1/(k^2+m_K^2)$
in the loop diagram, that contains the $s$-quark current coupling
(Fig.~1a), by the propagator multiplied by a dipole form factor 
\begin{equation}
v(k^2)\rightarrow \frac{1}{m_K^2+k^2}\Biggl[\frac{\Lambda^2-m_K^2}{
\Lambda^2+k^2}\Biggr]^2. \eqnum{2.7}
\end{equation}
Current conservation then demands that the product of the two meson
propagators in the loop amplitude that corresponds to the kaon current
loop (Fig.~1b) be modified as
\begin{equation}
\frac{1}{m_K^2+k_1^2}\frac{1}{m_K^2+k_2^2}\rightarrow
\frac{v(k_2^2)-v(k_1^2)}{k_1^2-k_2^2}. \eqnum{2.8}
\end{equation}
The calculation then proceeds by first calculating the strangeness
form factors $F_1^s$ and $F_2^s$ for the constituent
quarks. These form factors are the same for the $u$- and $d$-quarks. 
The relation of these form factors to the corresponding strangeness form
factors of the proton is simple only under the assumption that
$m_q=m_p/3$, which implies an equipartition of the total
proton momentum between the quarks:
\begin{equation}
F_1^s(Q^2)= 3F_{1q}^s(Q^2),\quad F_2^s(Q^2)={m_p\over m_q}F_{2q}
^s(Q^2).\eqnum{2.9}
\end{equation}
From these relations we obtain the charge and magnetic form factors of the
proton as
\begin{equation}
G_{E}^s(Q^2)= F_{1}^s(Q^2)-{Q^2\over 4 m_p^2}F_{2}^s(Q^2),\eqnum{2.10a}
\end{equation}
\begin{equation}
G_{M}^s(Q^2)=F_{1}^s(Q^2)+F_{2}^s(Q^2). \eqnum{2.10b}
\end{equation}
The magnetic form factor at $Q^2=0$ then yields the strangeness
magnetic moment in units of nuclear magnetons.

Charge conservation requires that $F_1^s(0)=0$. This requirement is
satisfied by subtracting the value of $F_1^s(0)$ from the
expression below.

The contribution to the strangeness form factor $F_1^s$ of the proton
from the two loop diagrams in Fig.~1 are obtained as
\begin{eqnarray}
F_{1q}^s&&(Q^2)\{a,K\}=\frac{g_{Kqs}^2}{8\pi^2}\int_{0}^{1}dx\,(1-x)
\int_{0}^{1}dy\, \Biggl\{\Bigl[(m_s-m_q)^2
+2m_q(m_s-m_q)(1-x)\nonumber\\
&&+m_q^2(1-x)^2-Q^2(1-x)^2(1-y)y\Bigr] K_1(Q^2)
+\ln\frac{H_1(\Lambda_\chi^2)}{H_1(m_K^2)}
-x\frac{\Lambda_\chi^2-m_K^2}{H_1(\Lambda^2)}\Biggr\} ,\eqnum{2.11a}\\
F_{1q}^s&&(Q^2)\{b,K\}=\frac{g_{Kqs}^2}{8\pi^2}\int_{0}^{1}dx\, x
\int_{0}^{1}dy\, \Biggl\{2m_q(m_s-m_qx)(1-x)K_2(Q^2)\nonumber\\
&&-\ln\frac{H_2(\Lambda_\chi^2)}{H_2(m_k^2)}+x\frac{\Lambda^2-m_k^2}{H_2(
\Lambda_\chi^2)}\Biggr\}.\eqnum{2.11b}
\end{eqnarray}
Here the functions $K_1(Q^2)$ and $K_2(Q^2)$ have been defined as
\begin{eqnarray}
K_1(Q^2)=&&\frac{1}{H_1(m_K^2)}-\frac{1}{H_1(\Lambda_\chi^2)}
-x\frac{\Lambda_\chi^2-m_K^2}{H_1^2(\Lambda_\chi^2)}
,\eqnum{2.12a}\\
K_2(Q^2)=&&\frac{1}{H_2(m_K^2)}-\frac{1}{H_2(\Lambda_\chi^2)}
-x\frac{\Lambda_\chi^2-m_K^2}{H_2^2(\Lambda_\chi^2)}.\eqnum{2.12b}
\end{eqnarray}
The denominator functions $H_1(m^2)$ and $H_2(m^2)$ are defined as
\begin{eqnarray}
H_1(m^2)=&&m_s^2(1-x)-m_q^2x(1-x)+m^2x
+Q^2(1-x)^2y(1-y),\eqnum{2.13a}\\
H_2(m^2)=&&m_s^2(1-x)-m_q^2x(1-x)+m^2x
+Q^2x^2y(1-y).\eqnum{2.13b}
\end{eqnarray}
After subtraction of the corresponding
values at $Q^2=0\,\,(F_1^s(0))$ from the form factors $F_1^s(Q^2)$ the
integrals remain finite even in the
limit $\Lambda_\chi^2\rightarrow \infty$.

The corresponding contributions to the strangeness Pauli form factors
of the quarks is
\begin{eqnarray}
F_{2q}^s(Q^2)\{a,K\}=&&-\frac{g^2}{4\pi^2}\int_{0}^{1}dx\,(1-x)^2
\int_{0}^{1}dy\,m_q(m_s-m_qx)K_1(Q^2),\eqnum{2.14a}\\
F_{2q}^s(Q^2)\{b,K\}=&&-\frac{g^2}{4\pi^2}\int_{0}^{1}
dx\,x(1-x)\int_{0}^{1}dy\, m_q(m_s-m_qx)K_2(Q^2).\eqnum{2.14b}
\end{eqnarray}
These expressions reduce to those given in Ref.~\cite{Hannelius} in the
limit $Q^2\rightarrow 0$, if multiplied by the factor $m_p/m_q$
to give strangeness magnetic moments in units of nuclear magnetons.

In Fig.~2 the kaon loop contributions to the proton strangeness 
Dirac form
factor $F^s_1$ are shown as functions of momentum transfer, after
subtraction of the irrelevant constant $F_1^s(0)$. This
loop contribution to the proton strangeness form factor is very small
and negative, and for $Q^2\leq 1$ (GeV/$c)^2$ it decreases slowly from
0 to $\sim -0.01$. As shown below, the magnitude of this contribution
is smaller than that of the strange vector meson loops.

The contributions from the kaon loop amplitudes to the strangeness
Pauli form factor $F_2^s$ are shown in Fig.~3. These
contributions, while small, are notably larger than the
corresponding vector meson loop contributions that are
derived in section 3 below.
The momentum dependence of the kaon loop contribution to
$F_2^s(Q^2)$ is fairly weak for $Q^2$ values below $1$ (GeV/c)$^2$.

\section{Strange vector meson loop fluctuations}

The coupling of $K^*$ mesons to constituent quarks is described by the
Lagrangian
\begin{equation}
{\cal L}_{K^*qs}=ig_{K^*qs}\bar\psi_s\Bigl(\gamma_\mu
+\frac{m_s-m_q}{m_{K^*}^2}\partial_\mu
+i\frac{\kappa_{K^*qs}}{m_s+m_q}\sigma_{\mu\nu}
\partial_\nu\Bigr)\sum_{a=4}^{7}\lambda^aK_\mu^a\psi
+\text{h.c.}.\eqnum{3.1}
\end{equation}
This coupling is a generalization to fermions of unequal mass of the
conventional transverse Proca coupling for vector mesons.

The coupling constants $g_{K^*qs}$ and $\kappa_{K^*qs}$ may be
determined from the corresponding couplings of $K^*$ mesons to the
baryon octet by the quark model relations~\cite{Hannelius}:
\begin{eqnarray}
g_{K^*qs}=&&g_{K^*\bar B B},\nonumber\\
g_{K^* qs}(1+\kappa_{K^*qs})=&&\frac{3}{5}\frac{m_s+m_q}{\bar M}
g_{K^*\bar BB}(1+\kappa_{K^*\bar BB}).\eqnum{3.2}
\end{eqnarray}
Here $\bar M$ represents the average of the nucleon and $S=-1$ hyperon
$(\Lambda,\Sigma)$ masses.

A recent comprehensive boson exchange potential model fit to
nucleon-nucleon scattering data gives $g_{K^*\bar BB}=2.97$ and
$\kappa_{K^*\bar BB}=4.22$, with a liberal uncertainty margin
\cite{Rijken}. These values yield $g_{K^*qs}^2/4\pi \simeq 0.7$ and
$\kappa_{K^*qs}\simeq 0.21$. The small value of the tensor coupling
$\kappa_{K^*qs}$ and its large uncertainty range suggests that it is
consistent with 0. At this stage it is therefore justified to neglect
the Pauli term in (3.1) altogether.

The current density operator for the $K^*$ mesons takes the form
\begin{equation}
j_\mu=\pm ie\{K_\nu^{*\dagger}\partial_\mu K_\nu^*
-K_\nu^{*\dagger}\partial_\nu K_\mu^*\}+{\rm h.c.} \eqnum{3.3}
\end{equation}
The contribution to the strangeness from factors $F_1^s$ of the $u$- and
$d$-quarks from the $K^*$ meson loops described by the Feynman diagrams
in Figs.~1a and b (when the meson line represents a $K^*$ meson)
are
\begin{eqnarray}
F_{1q}^s&&(Q^2)\{a,K^*\}=\frac{g_{K^*qs}^2}{4\pi^2}\int_{0}^{1}dx\,
(1-x)\int_{0}^{1}dy\, \Biggl\{\Bigl[m_s^2-4m_qm_s x
+m_q^2x^2\nonumber\\
&&-Q^2(x+(1-x)^2 y(1-y))\Bigr]\bar K_1(Q^2)
+\ln\frac{H_1(\Lambda_\chi^2)}{H_1(m^2_{K^*})}
-x\frac{\Lambda_\chi^2-m_{K^*}^2}{H_1(\Lambda_\chi^2)}\Biggr\}+{\cal O}
\Bigl(\frac{1}{m_{K^*}^2}\Bigr).\eqnum{3.4a}\\
F_{1q}^s&&(Q^2)\{b,K^*\}=\frac{g_{K^*qs}^2}{8\pi^2}\int_{0}^{1}dx\,
x\int_{0}^{1}dy\, \Biggl\{\Bigl[6m_q^2 x(1-x)-6m_q m_s(1-x)\nonumber\\
&&+Q^2 x(1-2xy(1-y))\Bigr]\bar K_2(Q^2)
+6\Bigl[ \ln\frac{H_2(\Lambda_\chi^2)}{H_2(m^2_{K^*})}
-x\frac{\Lambda_\chi^2-m_{K^*}^2}{H_2(\Lambda_\chi^2)}\Bigr] \Biggr\}
+{\cal O}\Bigl(\frac{1}{m_{K^*}^2}\Bigr).\eqnum{3.4b}
\end{eqnarray}
Here the auxiliary functions $\bar K_1(Q^2)$ and $\bar K_2(Q^2)$ have
been defined as the functions $K_1(Q^2)$ and $K_2(Q^2)$ in Eqs.~(2.12),
with the replacement of $m_K^2$ by $m_{K^*}^2$.

The terms of order $m_{K^*}^{-2}$ and higher powers of $m_{K^*}^{-2}$ in
(3.3) arise from the terms proportional to $m_{K^*}^2$ in the vector
meson propagator $(\delta_{\mu\nu}+k_\mu k_\nu/m_{K^*}^2)/(m_{K^*}^2+k^2)$
and the coupling (3.1). These terms are small in comparison to the terms
of Eqs.~(3.3) at low values of $Q^2$. The explicit expressions for the
contributions of order $m_{K^*}^{-2}$ to $F^s_{1q}$ that are indicated
in Eqs.~(3.4a) and (3.4b) from the two loop
diagram amplitudes illustrated in Figs.~1a and b are
\begin{eqnarray}
F_{1q}^s&&(Q^2)\{a,K^*,{\cal O}(m_{K^*}^{-2})\}
=\frac{g_{K^*qs}^2}{8\pi^2}\frac{m_q^2}{m_{K^*}^2}\int_{0}^{1}dx\,(1-x)
\int_{0}^{1}dy \nonumber\\ 
\Biggl\{\Bigl[
&&m_s^2(1-x)^2-{m_s^2\over m_q^2}Q^2(1-x)^2y(1-y)\nonumber\\
&&+m_sm_q[2x(1-x)+{Q^2\over m_q^2}(1-x)^2
(1+2xy(1-y))]+m_q^2x^2(1-x)^2\nonumber\\
&&+Q^2(1-x)^2(2x^2y(1-y)-y(1-y^2)+x(1-3y+4y^2-y^3))\nonumber\\
&&+\frac{Q^4}{m_q^2}(1-x)^2(1-y)y(x+(1-x)^2(1-y)y)\Bigr]\bar K_1(Q^2)
+\Bigl[{m_s^2\over m_q^2}+2{m_s\over m_q}(1-3x)\nonumber\\
&&+1+6x(1-x)+\frac{Q^2}{m_q^2}(2-3x-6(1-x)^2y(1-y))\Bigr]
\Bigl [\ln\frac{H_1(\Lambda_\chi^2)}{H_1(m_{K^*}^2)}
-x\frac{\Lambda_\chi^2-m_{K^*}^2}{H_1(\Lambda_\chi^2)}\Bigr ]\nonumber\\
&&+\frac{6}{m_q^2}\Bigl [H_1(m_{K^*}^2)\ln\frac{H_1(m_{K^*}^2)}{
H_1(\Lambda_\chi^2)}
-H_1(m_{K^*}^2)+H_1(\Lambda_\chi^2)\Bigr ]\Biggr\},\eqnum{3.5a}\\
F_{1q}^s&&(Q^2)\{b,K^*,{\cal O}(m_{K^*}^{-2})\}=
{g_{K^*qs}^2\over 8\pi^2}{m_q^2\over m_{K^*}^2}
\int dx\,x\int dy\nonumber\\
&&\Biggl\{\Bigl [-m_sm_q[(1-x)^3+{Q^2\over m_q^2}(1-x)]\Bigr.
+Q^2x^2(1-x)[1-2y(1-y)(1+x)]\nonumber\\
&&+{Q^4\over m_q^2}x^2y(1-y)[1-x+2x^2y(1-y)]\Bigr ]\bar K_2(Q^2)\nonumber\\
&&+\Bigl[6{m_s\over m_q}(1-x)+6(1-x)^2-(1-x)(3-5x)\Bigr.\nonumber\\
&&+{Q^2\over m_q^2}(-2+3x-11x^2y(1-y))\Bigr]
\Bigl[\ln{H_2(\Lambda_\chi^2)\over
H_2(m_{K^*}^2)}-x{\Lambda_\chi^2-m_{K^*}^2\over H_2(\Lambda_\chi^2)}
\Bigr]\nonumber\\
&&+{9\over m_q^2}\Bigl[H_2(m_{K^*}^2)\ln{H_2(m_{K^*}^2)\over H_2(
\Lambda_\chi^2)}-H_2(m_{K^*}^2)+H_2(\Lambda_\chi^2)\Bigr]\Biggr\}
\eqnum{3.5b}
\end{eqnarray}
The contributions from the $K^*$ loop diagrams in Fig.~1 to the
strangeness factors $F_2^s$ of the $u$- and $d$-quarks are obtained as
\begin{eqnarray}
F_{2q}^2&&(Q^2)\{a,K^*\}=\frac{g_{K^* qs}^2}{4\pi^2}\int_{0}^{1}dx\,
(1-x)\int_{0}^{1}dy\,
2m_q x[2(m_s-m_q)+m_q(1-x)]\bar K_1(Q^2)\nonumber\\
&&+{\cal O} \Bigl(\frac{1}{m_{K^*}^2}\Bigr),
\eqnum{3.6a}\\
F_{2q}^s&&(Q^2)\{b,K^*\}=\frac{g_{K^*qs}^2}{4\pi^2}\int_{0}^{1}dx\,
x\int_{0}^{1}dy\,
m_q[m_s(3x-2)-m_qx(2x-1)]\bar K_2(Q^2)\nonumber\\
&&+{\cal O}(\frac{1}{m_{K^*}^2}).
\eqnum{3.6b}
\end{eqnarray}
Finally the corresponding terms that are proportional to $1/m_K^{*2}$
have the expressions:
\begin{eqnarray}
F_{2q}^2&&(Q^2)\{a,K^*,{\cal O}(m_{K^*}^{-2})\}=
-\frac{g_{K^* qs}^2}{4\pi^2}\frac{m_q^2}{m_{K^*}^2}
\int_{0}^{1}dx\,(1-x)\int_{0}^{1}dy\,\Biggl\{
\Bigr[(1-x)^2(m_s-m_q)(m_s+m_qx)\nonumber\\
&&+\frac{Q^2}{m_q}(m_s-m_q)(1-x^2)(1-x)(1-y)y
\Bigl]\bar K_1(Q^2)\nonumber\\
&&-2(1-\frac{m_s}{m_q})(1-\frac{3}{2}x)
(\ln \frac{H_1(\Lambda_\chi^2)}{H_1(m_{K^*}^2)}-
x\frac{\Lambda_\chi^2-m_{K^*}^2}{H_1(\Lambda_\chi^2)})
\Biggr\}
\eqnum{3.7a}\\
F_{2q}^s&&(Q^2)\{b,K^*,{\cal O}(m_{K^*}^{-2})\}=
\frac{g_{K^* qs}^2}{4\pi^2}\frac{m_q^2}{m_{K^*}^2}
\int_{0}^{1}dx\,x \int_{0}^{1}dy\,\Biggl\{\Bigl [
m_q(1-x)^2(m_s-m_qx^2)\nonumber\\
&&+Q^2x^2y(1-y)(2-2x+x^2-\frac{m_s}{m_q})
\Bigr ]\bar K_2(Q^2)\nonumber\\
&&-2\Bigl[\frac{m_s}{m_q}-2(1-x)^2-x\Bigr]
\Bigl[\ln\frac{H_2(\Lambda_\chi^2)}
{H_2(m_{K^*}^2)}-x\frac{\Lambda_\chi^2-m_{K^*}^2}
{H_2(\Lambda_\chi^2)}\Bigr]
\Biggr\}
\eqnum{3.7b}
\end{eqnarray}
These expressions reduce to those derived in Ref.~\cite{Hannelius} in
the limit $Q^2\rightarrow 0$, once multiplied by $m_p/m_q$ in order to
obtain the results in units of nuclear magnetons.

The strange vector meson loop contributions to the strangeness
Dirac form factor $F_1^s(Q^2)$ of the proton are obtained
after subtraction of the irrelevant values $F_{1q}^s(0)$
from the expressions (3.4) and (3.5) and multiplication of
the sum of the remainders by a factor 3 (2.9). This
contribution is shown in Fig.~2 along with the corresponding
kaon loop contribution. In this case the vector meson
loop contribution is larger in magnitude than the kaon
loop contribution. Even so the sum of the $K$ and $K^*$
loop contributions remains very small in magnitude, reaching
only the value -0.086 around $Q^2=1~($GeV/c$)^2$.

The contribution of the strange vector meson loops to the strangeness
Pauli form factor is very small because of a near cancellation between
the two involved loop diagrams~\cite{Hannelius}. This contribution is shown
in Fig.~3, with the much larger contribution from the kaon loop
diagrams.

\section{The $K^*-K$ loop contribution}

We finally consider the strangeness loop fluctuation, for which the
e.m. coupling is to the $K^*K$ transition vertex (Fig.~4). The amplitude of
this loop fluctuation may be calculated from the empirically known
radiative widths of the $K^*$ mesons. The $K^*K$ transition current
vertex has the form
\begin{equation}
\langle K^a(k')|J_\mu|K^{*b}_\sigma (k)\rangle
= -i\frac{g_{{K^*}K\gamma}}{m_{K^*}}
\epsilon_{\mu\lambda\nu\sigma} k_\lambda k'_\nu
\delta^{ab}\eqnum{4.1}
\end{equation}
The coupling constant $g_{{K^*}K\gamma}$ depends on the charge state of
the strange mesons. It may be determined from the radiative decay widths
using the expression~\cite{Hannelius}
\begin{equation}
\Gamma (K^* \to K\gamma)=\alpha\frac{g_{{K^*}K\gamma}^2}{24\pi}
m_{K^*} \Bigl[1-\Bigl(\frac{m_K}{m_{K^*}}\Bigr)^2\Bigr]^3.\eqnum{4.2}
\end{equation}
Here $\alpha$ is the fine structure constant.

Given the empirical radiative widths $\Gamma ({K^*}^+ \to K^+\gamma)=50$ keV
and $\Gamma ({K^*}^0 \to K^0\gamma)=116$ keV~\cite{PDG}, the
corresponding coupling constant values are obtained as $g_{{K^*}^\pm
K^\pm \gamma}=0.75$ and $g_{{K^*}^0 K^0 \gamma}=1.14$ (with the sign
convention of~\cite{Hannelius}).

The $K^*K$ loop diagrams (Fig.~4) only contribute to the strange Pauli
form factors of the quarks. With the convention of assigning the $K$ and
$K^*$ mesons a ``strangeness charge'' of $-1$, the contribution to the
strangeness Pauli form factor $F_{2q}^s(Q^2)$ from these loop diagrams
is found to be
\begin{eqnarray}
F_{2q}^s (Q^2)=&&-\frac{g_{Kqs}g_{K^*qs}g_{K^*K\gamma}}{2\pi^2}
\frac{m_q}{m_{K^*}}\int_0^1 dx\, x \int_0^1 dy \Biggl\{
m_q(1-x)(m_s-m_q x)\nonumber\\
&&\times (\frac{1}{G_1}-\frac{1}{G_2}-\frac{1}{G_3}
+\frac{1}{G_4})-\ln\Biggl(\frac{G_2 G_3}{G_1 G_4}\Biggr)\Biggr\}.
\eqnum{4.3}
\end{eqnarray}
The quantities $G_i$ here have been defined as
\begin{equation}
\begin{array}{lll}
G_1=G(m_K,m_{K^*}),\quad & G_2=G(\Lambda_\chi,m_{K^*}),\\
G_3=G(m_K,\Lambda_\chi),\quad & G_4=G(\Lambda_\chi,\Lambda_\chi).
\end{array}
\eqnum{4.4}
\end{equation}
The auxiliary function $G(m,m')$ is defined as
\begin{equation}
G(m,m')=m_s^2 (1-x)-m_q^2 x(1-x)+m^2 x(1-y)
+m'^2 xy + Q^2x^2 y(1-y).
\eqnum{4.5}
\end{equation}
These expressions represent direct generalizations of the corresponding
expressions defined in Ref.~\cite{Hannelius} for the case $Q^2=0$.

To account for the different coupling constants $g_{K^*K\gamma}$ in
the case of $u$- and $d$-quarks, (2.9) is generalized to
\begin{equation}
F_2^s(Q^2)=\frac{m_p}{m_s}\Biggl [ \frac{4}{3} F_{2u}^s(Q^2)
-\frac{1}{3} F_{2d}^s(Q^2)\Biggr ].\eqnum{4.6}
\end{equation}

The $K^*K$ loop contribution to $F_2^s (Q^2)$ is shown in Fig.~3.
This loop contribution has the opposite sign to that of the diagonal
strangeness loop fluctuations.

\section{SAMPLE and HAPPEX}

The SAMPLE \cite{Sample1,Sample2} experiment measures the strangeness 
magnetic form
factor of the proton $G_M^s$ at $Q^2=0.1$ (GeV/$c)^2$. The HAPPEX
\cite{Happex} experiment measures the combination $G_E^s+0.39G_M^s$ at
$Q^2=0.48$ (GeV/$c)^2$. The contributions to these observables from the
kaon and $K^*$ loops in Figs.~1 and 4 are shown in Figs.~5 and 6 as a
function of momentum transfer.

The cut-off dependence is small~\cite{Hannelius}, and
for $\Lambda_\chi=1.2$ GeV we
obtain $G_M^s(0.1)=-0.06$ and $G_E^s(0.48) +0.39 G_M^s(0.48)=-0.08$.
The latter value is slightly below the uncertainty range of the result of
the HAPPEX experiment. Similarly the former value is within the
uncertainty range of the result of the SAMPLE experiment, provided that
$G_A^Z$ is small and positive rather than large and negative as originally
suggested \cite{Musolf,Musolf91}.

The smallness of the calculated strangeness observables of the proton is an
inherent feature of the chiral quark model. The smallness of the
measured strangeness observables suggests that this model may provide
a useful framework for describing those observables. The calculated
momentum dependence of $G_E^s$ agrees fairly well with that obtained
by heavy baryon chiral perturbation theory \cite{Hemmert}. The
strangeness radius is $\sim 0.02$ fm$^2$, which value is also obtained in the
baryon loop calculations in Ref.~\cite{Barz2} if the cut-off scale is
set to the chiral symmetry restoration scale. Because of the large
negative vector meson contribution to the Dirac form factor $F_1^s$ the 
calculated $G_M^s$ form factor grows more negative with increasing
$Q^2$, whereas the third order chiral perturbation theory result, which
does not consider vector mesons, for $G_M^s$ is that it increases
slowly with momentum transfer.
Hitherto QCD lattice calculations have been made only for the
strangeness magnetic moment $G_M^s(0)$, but not for the form factors.
The calculated values are negative, with substantial uncertainty
limits (-0.36 $\pm$ 0.20 \cite{liu}, -0.16 $\pm$ 0.18 \cite{lein}).
The chiral quark model value for $G_M^s(0)\sim -0.06$ 
\cite{Hannelius} falls within
the uncertainty range of the latter value.

\acknowledgments
We are grateful for the hospitality of Professor R. D. McKeown at the
W. K. Kellogg Radiation Laboratory of the California Institute of
Technology where this work was completed. L. H. thanks the Waldemar
von Frenckell foundation for a stipend. This work was supported in
part by the Academy of Finland under contract 43982.

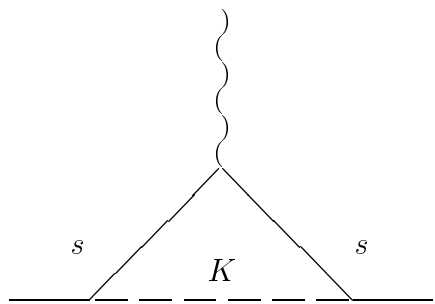
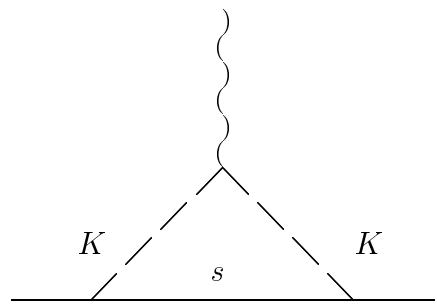
\begin{figure}
\caption{Kaon and $K^*$ loop fluctuations of constituent quarks, 
which contribute to the strangeness form factors of the proton.}
\end{figure}

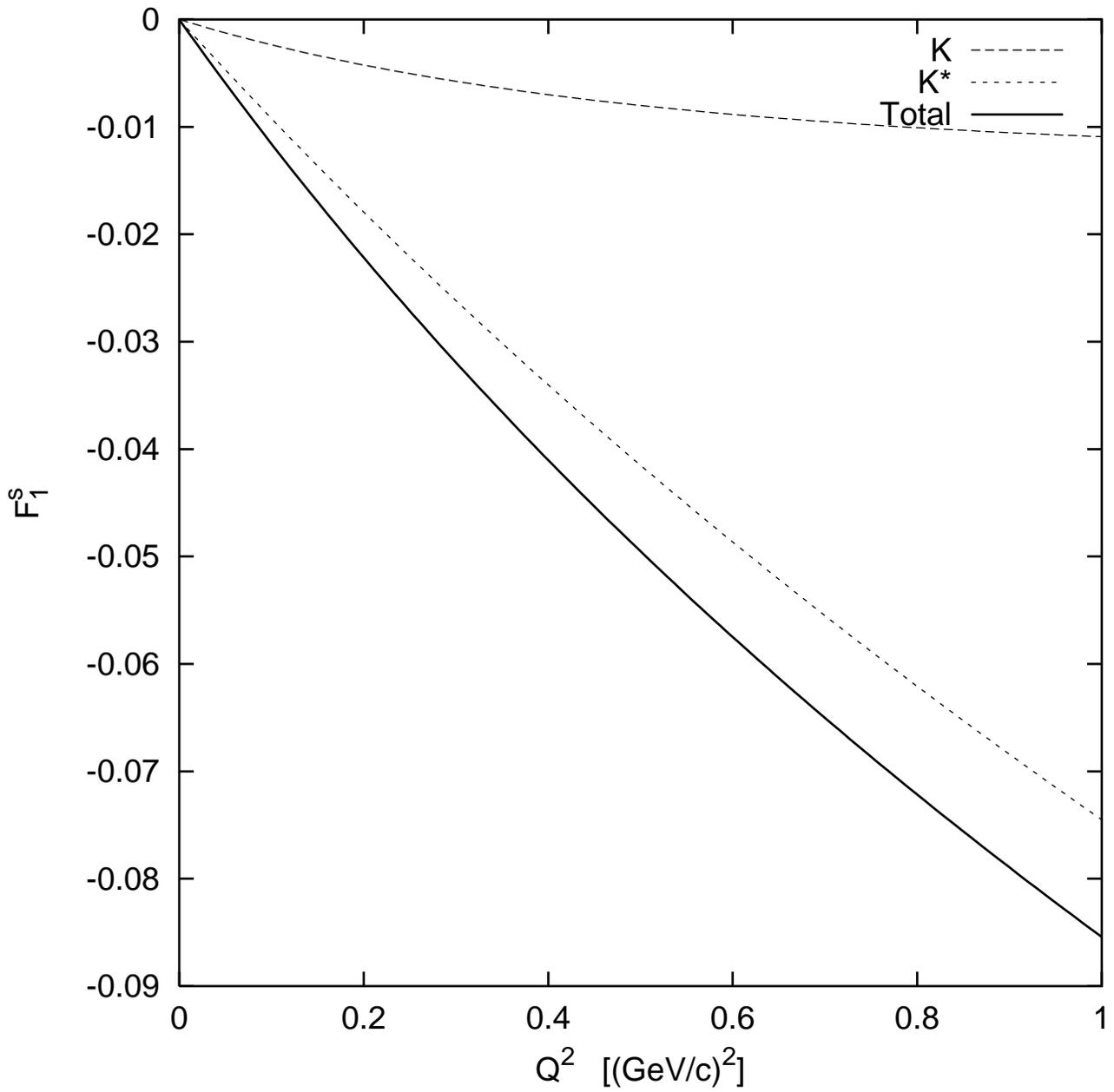
\begin{figure}
\caption{The Dirac strangeness form factor $F_1^s$ of the proton as a
function of momentum transfer.}
\end{figure}

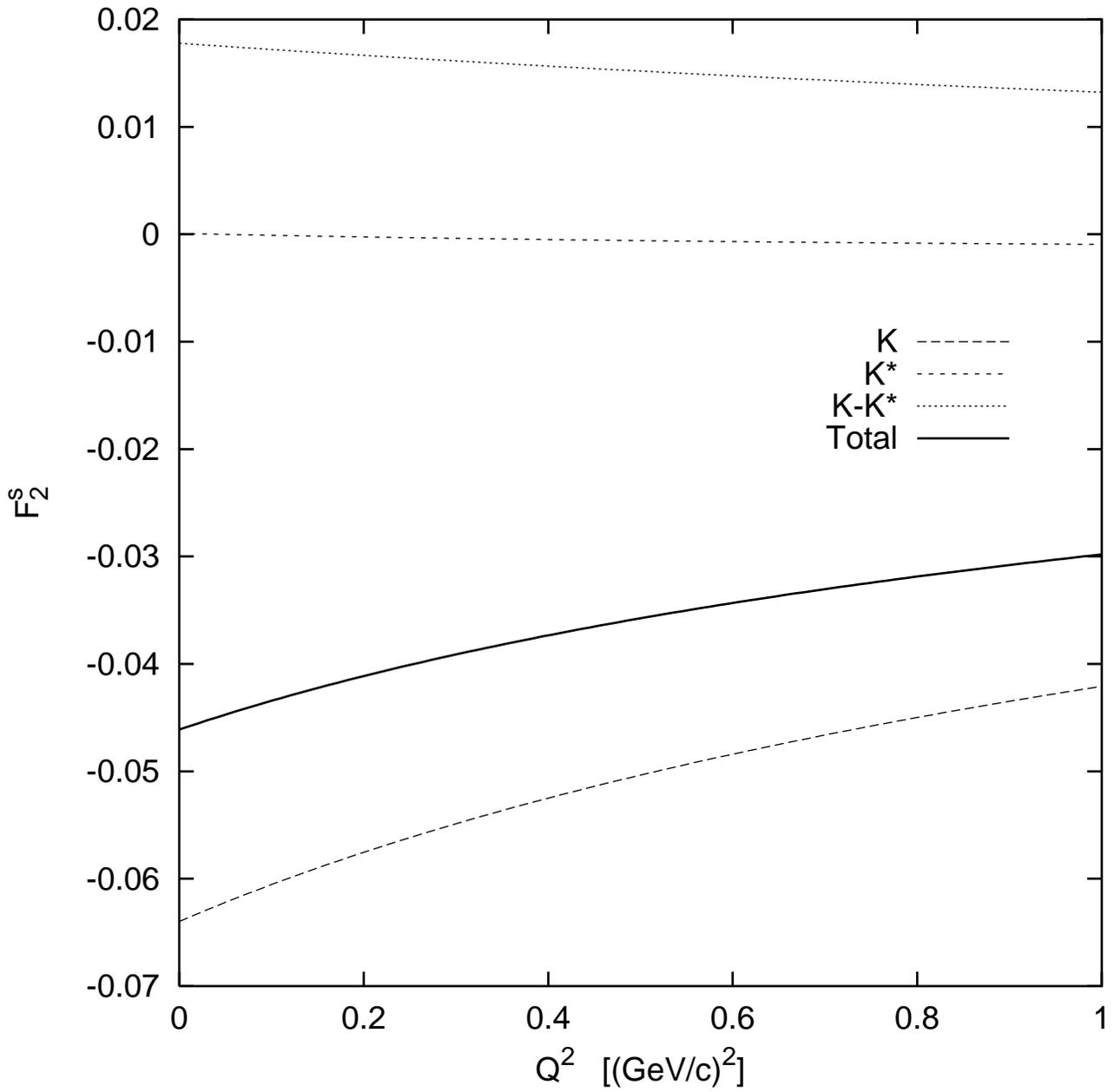
\begin{figure}
\caption{The Pauli strangeness form factor $F_2^s$ of the proton as a
function of momentum transfer.}
\end{figure}

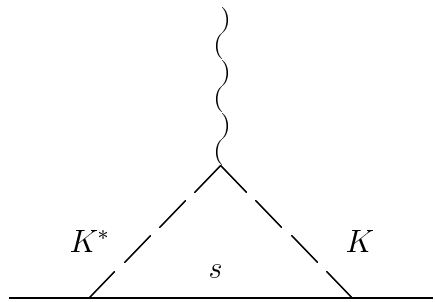
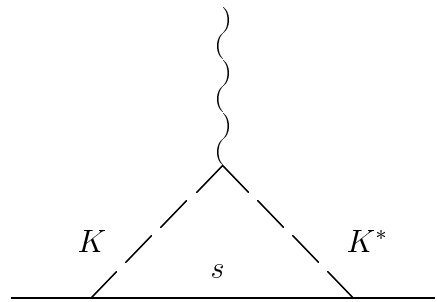
\begin{figure}
\caption{Strangeness fluctuations of the constituent quarks,
which involve an intermediate radiative $K^*\rightarrow K\gamma$
transition.}
\end{figure}

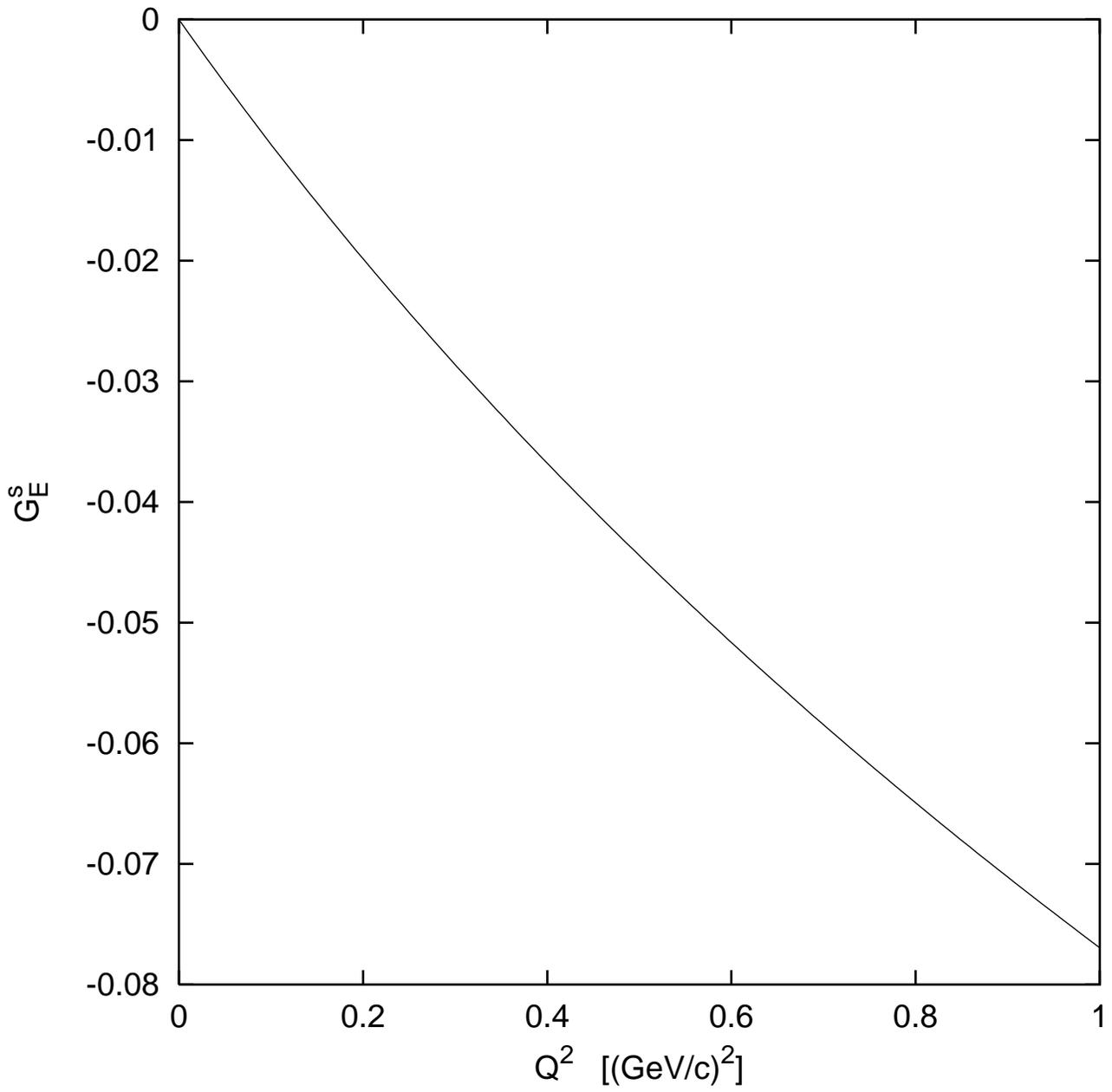
\begin{figure}
\caption{The Sach's strangeness form factor $G_E^s$ of the proton as a
function of momentum transfer.}
\end{figure}

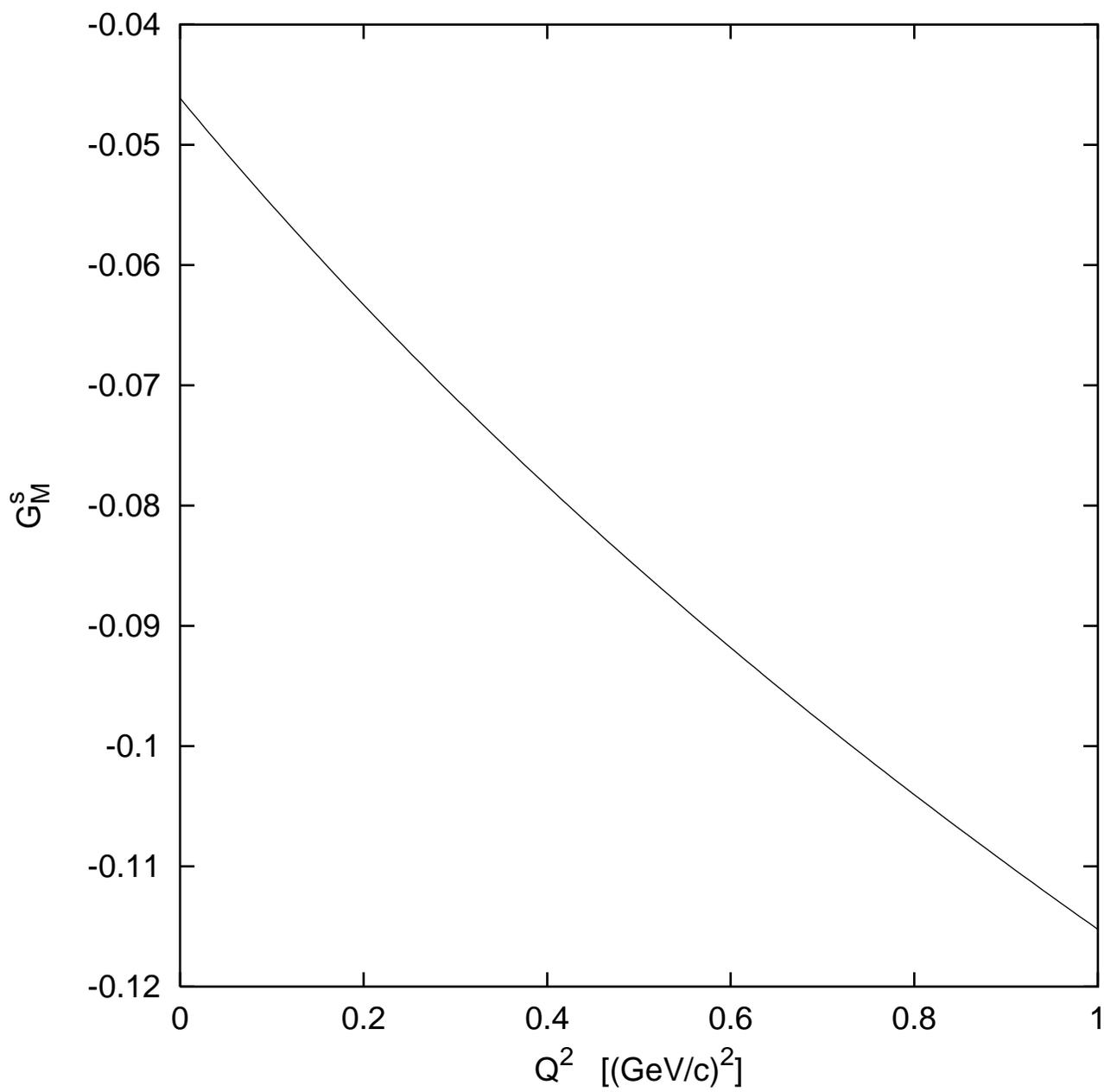
\begin{figure}
\caption{The Sach's strangeness magnetic form factor $G_M^s$ of the
proton as a function of momentum transfer. The SAMPLE
experiment gives the experimental value at $Q^2=0.1$ (GeV/c)$^2$.}
\end{figure}

\end{document}